\documentclass[prl,amsmath,amssymb,twocolumn,superscriptaddress]{revtex4}
\usepackage{color}

\usepackage{graphicx}
\usepackage{latexsym}
\usepackage{dcolumn}

\usepackage{bm}
\usepackage{bibentry} 
\usepackage{bm}
\begin{document}

\preprint{APS/123-QED}

\title{Phonon-Dressed Two-Dimensional Carriers on the ZnO Surface}

\author{R. Yukawa}
\affiliation{Institute for Solid State Physics, The University of Tokyo, Kashiwa, Chiba 277-8581, Japan}
\author{K. Ozawa} 
\affiliation{Department of Chemistry, Tokyo Institute of Technology, Meguro, Tokyo 152-8551, Japan}
\author{S. Yamamoto}
\affiliation{Institute for Solid State Physics, The University of Tokyo, Kashiwa, Chiba 277-8581, Japan}
\author{H. Iwasawa}
\affiliation{Hiroshima Synchrotron Radiation Center, Hiroshima University, Higashi-Hiroshima, Hiroshima 739-0046, Japan}
\author{K. Shimada}
\affiliation{Hiroshima Synchrotron Radiation Center, Hiroshima University, Higashi-Hiroshima, Hiroshima 739-0046, Japan}
\author{E. F. Schwier}
\affiliation{Hiroshima Synchrotron Radiation Center, Hiroshima University, Higashi-Hiroshima, Hiroshima 739-0046, Japan}
\author{K. Yoshimatsu}
\affiliation{Institute of Materials Structure Science, High Energy Accelerator Research Organization (KEK), Tsukuba 305-0801, Japan}
\author{H. Kumigashira}
\affiliation{Institute of Materials Structure Science, High Energy Accelerator Research Organization (KEK), Tsukuba 305-0801, Japan}
\author{H. Namatame}
\affiliation{Hiroshima Synchrotron Radiation Center, Hiroshima University, Higashi-Hiroshima, Hiroshima 739-0046, Japan}
\author{M. Taniguchi}
\affiliation{Hiroshima Synchrotron Radiation Center, Hiroshima University, Higashi-Hiroshima, Hiroshima 739-0046, Japan}
\author{I. Matsuda}
\affiliation{Institute for Solid State Physics, The University of Tokyo, Kashiwa, Chiba 277-8581, Japan}

\email{imatsuda@issp.u-tokyo.ac.jp}

\date{\today}


\begin{abstract}
Two-dimensional (2D) metallic states formed on the ZnO(10$\bar{1}$0) surface by hydrogen adsorption have been investigated using angle-resolved photoelectron spectroscopy (ARPES). The observed metallic state is characterized by a peak-dip-hump structure at just below the Fermi level and a long tail structure extending up to 600 meV in binding energy. The peak and hump positions are separated by about 70 meV, a value close to the excitation energy of longitudinal optical (LO) phonons. Spectral functions formulated on the basis of the 2D electron-phonon coupling well reproduce the ARPES intensity distribution of the metallic states. This spectral analysis suggests that the 2D electrons accumulated on the ZnO surface couple to the LO phonons and that this coupling is the origin of the anomalous long tail. Our results indicate that the 2D electrons at the ZnO surface are described as the electron liquid model.

\end{abstract}

\maketitle
Two-dimensional electron systems (2DESs) in oxide semiconductors have attracted growing interests since the discovery of the high mobility electron gas at the LaAlO$_3$/SrTiO$_3$ heterojunction \cite{Ohtomo2004}. A series of angle-resolved photoelectron spectroscopy (ARPES) studies have proved that such 2DESs with metallic properties are also realized on various metal oxide surfaces such as ZnO \cite{Piper2010,Ozawa2010,Ozawa2010a,Ozawa2011}, In$_{2}$O$_{3}$ \cite{Zhang2013}, CdO \cite{Piper2008,King2010}, SrTiO$_3$ \cite{Meevasana2011,Santander-Syro2011,DAngelo2012,Yukawa2013,Wang2016,Rodel2016}, KTaO$_3$ \cite{King2012}, BaTiO$_3$ \cite{Rodel2016}, and anatase TiO$_2$ \cite{Rodel2015,Rodel2016} by chemical or physical doping of electrons to these surfaces. Based on the orbital characters of the metallic bands, the 2DESs are classified into two types; one is the $s$-orbital type like ZnO, In$_2$O$_3$, and CdO and the other is the $d$-orbital type like SrTiO$_3$, KTaO$_3$, BaTiO$_3$, and anatase TiO$_2$. 

A characteristic difference among the 2DESs with different orbital character is the many-body effects of the electronic system such as electron-phonon ($e$-ph) and electron-plasmon interactions. Recent ARPES studies have reported that the many-body interactions in the 2DESs on SrTiO$_3$(001) are inevitable and the 2DESs should be described in terms of the electron liquid \cite{Meevasana2011,Wang2016}.
In these 2DESs, the tails that accompany the two-dimensional (2D) metallic band have been attributed to energy loss structures by the $e$-ph coupling interaction.  A similar enhanced spectral weight has also been observed for the 2DESs on anatase TiO$_2$ surfaces \cite{Rodel2015}.  On the other hand, no such anomalous spectral weight has been reported for the $s$-orbital derived 2D metallic bands on ZnO(10$\bar{1}$0) \cite{Ozawa2011,Ozawa2010,Ozawa2010a}, ZnO(000$\bar{1}$) \cite{Piper2010,Ozawa2011}, In$_{2}$O$_{3}$(111) \cite{Zhang2013}, and CdO(001) surfaces \cite{Piper2008,King2010}. This systematic difference implies that the $e$-ph coupling strength might be weaker for the $s$ electrons than the $d$ electrons. However, this difference is against the theoretical predictions \cite{Frohlich1950} that the orbital character should have little effect on the $e$-ph coupling. Therefore, the solution of this discrepancy is one of the challenging subjects for the 2DESs on oxide surfaces.

In the present study, the electronic structure of the 2DES developed on the H-dosed ZnO(10$\bar{1}$0) surface was examined by ARPES. The detailed measurements have allowed us to successfully extract 2D quasiparticle (2DQP) bands and the $e$-ph satellites from the measured broad band, which has been regarded as a single metallic band in the previous studies \cite{Piper2010,Ozawa2011,Ozawa2010,Ozawa2010a,Deinert2015}. The spectral analysis reveals the $e$-ph coupling strength of $\alpha= 0.30-0.34$, indicating a sufficient $e$-ph coupling in the 2DES on the ZnO(10$\bar{1}$0) surface.

Single-crystal ZnO(10$\bar{1}$0) was cleaned $in$ $situ$ by well-established procedures \cite{Ozawa2007,sup}. H$_2$ molecules were cracked by hot tungsten filaments and dosed on the surface at room temperature. By these procedures, surface-localized O 2$p$ dangling-bond state is suppressed and a 2DES is induced at the surface \cite{Ozawa2010a}.
The ARPES measurements were performed at BL-1 of a compact electron-storage ring (HiSOR) at Hiroshima University \cite{SHIMADA2002,Hayashi2013,sup}.

Figure 1(a) shows an ARPES intensity map of the 2D metallic band on the H-dosed ZnO(10$\bar{1}$0) surface.  The metallic band is only observable around the $\bar{\Gamma}$  point [Fig. 1(b)].  The band crosses the Fermi level ($E_{\text{F}}$) and forms a circular Fermi surface with a radius of a Fermi wave vector $k_{\text{F}} = 0.77$ nm$^{-1}$, which is determined from the peak position of the momentum distribution curve (MDC) at $E_{\text{F}}$ [Fig. 1(c)].  From the size of the Fermi surface, the density of the doped electrons at the surface is evaluated to be $9.4 \times 10^{12}$   cm$^{-2}$. This carrier concentration is well within a metallic regime at this surface as judged by the Mott-Ioffe-Regel criterion \cite{sup}.

\begin{figure}
\centering
\includegraphics[width=7.5cm]{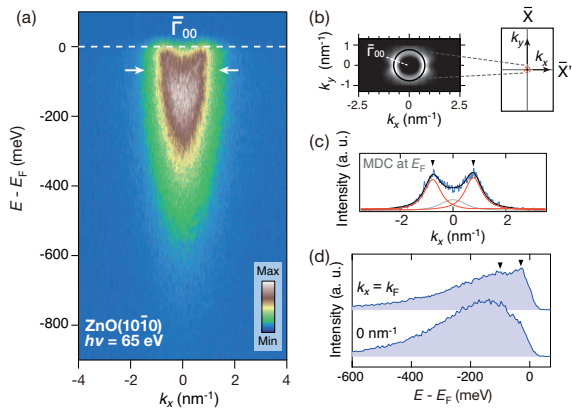}
\caption{(color online). Results of the ARPES measurements for the H-dosed ZnO(10$\bar{1}$0) surface with $h\nu = 65$ eV at 13 K.  (a) An intensity map near the $\bar{\Gamma}_{00}$ point along the $\bar{\Gamma}-\bar{\text{X'}}$ axis.  Arrows indicate the intensity dips at $E_{\text{F}}-70$ meV.  (b) The curvature intensity plot \cite{Zhang2011} of the Fermi surface.  The surface Brillouin zone is also shown. (c) An MDC of the 2D metallic state at $E_{\text{F}}$.  The experimental lineshape was fitted by three Voigt functions: two represent the 2DQP states and one represents the background contribution. The positions of the 2DQP states at $E_{\text{F}}$ are indicated by triangles. (d) The EDCs at $k_x = 0$ nm$^{-1}$ and $k_{\text{F}}$. The peak and hump structures of the EDC at $k_{\text{F}}$ are indicated by triangles.
}
\label{fig1}
\end{figure} 

As previously reported \cite{Piper2010,Ozawa2011,Ozawa2010,Ozawa2010a,Deinert2015}, the 2D metallic band has a very broad spectral feature with a long tail extending to 600 meV below $E_{\text{F}}$ at $k_x =0$ nm$^{-1}$ [Fig. 1(a)], where $k_x$ is a wave vector along the $\bar{\Gamma}-\bar{\text{X'}}$  axis [Fig. 1(b)].  A close examination of the metallic band, however, reveals that there are faint intensity dips at $E_{\text{F}} - 70$ meV.   The multiple-component structure is more obvious when we examine an energy distribution curve (EDC) at each $k_x$ point [Fig. 1(d)].  The lineshape of each EDC shows a fine structure at $<E_{\text{F}} - 100$ meV, indicating the contribution of at least two components.  The EDC at $k_{\text{F}}$ has a peak-dip-hump structure, which is similar to those observed for strongly correlated systems \cite{Norman1998,Cuk2005}. The energy difference between the peak and the hump in these systems roughly reflects energies of many-body interactions such as $e$-ph and electron-plasmon interactions \cite{Moser2013}. In the case of the present ZnO system, the energy difference between the peak and the hump is around 70 meV, which is close to the energy of the longitudinal optical (LO) phonon of the ZnO crystal ($\hbar\omega_{\text{ph}}=72$ meV \cite{Butkhuzi1998}).  Thus, we speculate that the $e$-ph coupling should be responsible for the multiple-component structures; namely, the peak and the hump originate from the 2DQP states and a phonon satellite (a 2D $e$-ph satellite), respectively. A possible contribution of the electron-plasmon coupling to the hump peak is excluded since the plasmon excitation energy at the density of  $9.4 \times 10^{12}$ cm$^{-2}$ exceeds 100 meV \cite{Goldstein1980}.  

In order to extract the intrinsic 2DQP band, the peak positions in the MDC curves were determined between $E_{\text{F}}+10$ meV and $E_{\text{F}}-20$ meV, where the two-peak structure as shown in Fig. 1(c) is obvious. The peak positions, plotted by circles in Fig. 2(a), were fitted by a parabola to reproduce the overall band structure. The fitting result gives a band bottom of $E_{\text{F}}-86$ meV at the $\bar{\Gamma}$ point and an effective mass of 0.26$m_e$, where $m_e$ is the mass of a free electron. Since this band naturally corresponds to the first subband of the 2D electrons at the ZnO surface, a depth profile of the potential well and the electron density as well as the energy position of a second subband can be deduced by self-consistently solving the Poisson-Schr{\"{o}}dinger equation \cite{Eger1979,Yukawa2015}. The results are shown in Fig. 2(b). The accumulated electrons are confined in a deep potential well formed between the vacuum and the edge of the Zn 4$s$-derived conduction band, i.e., the conduction band minimum (CBM).  The CBM at the surface is located at $E_{\text{F}}-478$ meV.  The energy minima of the first subband ($E_1$) and the second subband ($E_2$) are, respectively, $E_1 = E_{\text{F}} - 86$ meV (the experimental value) and $E_2 = E_{\text{F}} - 3$ meV at the $\bar{\Gamma}$ point. Because of a much lower density of electrons near the surface in the second subband in comparison with the first one [see inset of Fig. 2(b)], a contribution to ARPES intensities from the second subband is negligible but may be visible near $k_x =0$ nm$^{-1}$ as a background [Figs. 1(a,c)].

\begin{figure}
\centering
\includegraphics[width=7.5cm]{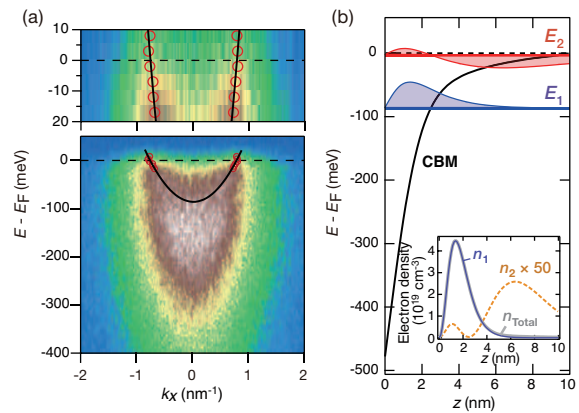}
\caption{ (color online). (a) Expanded images of Fig. 1(a). The positions of the 2DQP states are indicated by circles. The fitted result of the parabolic curve is drawn by solid lines. (b) Depth profiles of the CBM position. The corresponding eigenfunctions of the first and second subbands are displayed. The depth profile of the carrier electron density is shown in the inset.
}
\label{fig2}
\end{figure}

An important conclusion drawn from the calculated potential structure is that the long tail of the 2DQP band shown in Fig. 1(a) should not be ascribed to any electronic band because the tail extends into the band-gap region below the CBM.  This supports our claim that the broad feature of the band should arise from the $e$-ph interaction. 

A recent ARPES study on the metallized SrTiO$_3$(001) surface by UV irradiation has revealed that a long tail accompanying the 2DQP band can be described by the phonon satellite peaks \cite{Wang2016}.
 Here we show that the phonon satellite peaks also emerge at the higher-binding-energy side of the 2DQP band on the H-dosed SrTiO$_3$(001) surface, where adsorbed H donates its 1$s$ electron to the surface and induces Ti 3$d$-derived metallic bands \cite{DAngelo2012,Yukawa2013,sup}. Figure 3(a) shows an ARPES intensity map of the metallic band, whose tail extends to $E_{\text{F}}-500$ meV. An in-gap state is seen at $> E_{\text{F}}-500$ meV with a non-dispersive feature, in good agreement with the previous ARPES studies \cite{DAngelo2012,Yukawa2013}. The EDCs at $k_x = 0$ nm$^{-1}$ and $k_{\text{F}}$ exhibit a peak-dip-hump structure [Fig. 3(b)], which resembles the EDC at $k_{\text{F}}$ on the H-dosed ZnO surface [Fig. 1(d)].  The overall lineshape of each EDC is reproduced by two peaks labeled $A^{(0)}$ and  $A'^{(0)}$, which are associated with the first and second Ti 3$d_{xy}$-derived subbands, respectively, and three energy-loss satellite peaks labeled  $A^{(1)}$,  $A^{(2)}$, and  $A^{(3)}$, which are formed via the excitations of the LO phonon mode of SrTiO$_3$. 

\begin{figure}
\centering
\includegraphics[width=8cm]{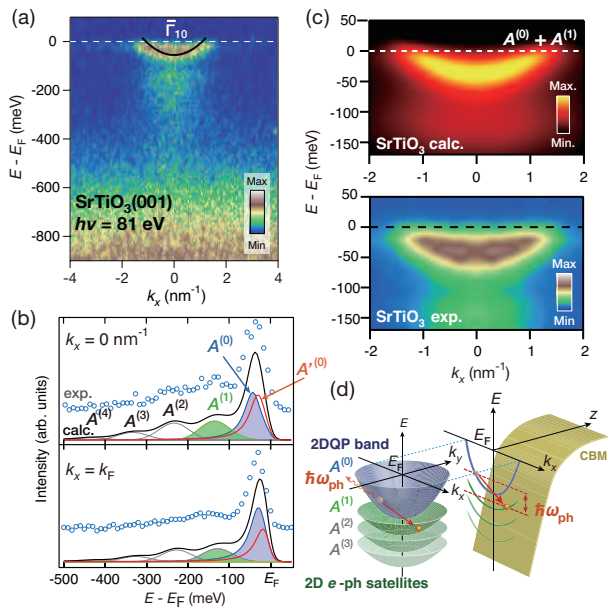}
\caption{ (color online). (a) The ARPES intensity map near the $\bar{\Gamma}$ point in the second surface Brillouin zone ($\bar{\Gamma}_{10}$) of the H-dosed SrTiO$_3$(001) surface ($h\nu = 81$ eV, 20 K). (b) The EDCs (circles) and calculated spectral functions (solid curves) at $k_x = 0$ nm$^{-1}$ and $k_{\text{F}}$. (c) The ARPES intensity map (a lower panel), which is symmetrized with respect to the center line ($k_x = 0$ nm$^{-1}$) and smoothed, is compared with the calculated spectral functions of $A^{(0)} + A^{(1)}$ (upper). (d) Schematic image of the 2D $e$-ph coupling. The brownish curve indicates the band bending structure near the surface. A blue curve indicates the 2DQP band, while green to pale-green curves represent the 2D $e$-ph satellites. The scattering process of an electron is illustrated by yellow circles.
}
\label{fig3}
\end{figure}

The  $A^{(0)}$ and $A^{(1)}$  peaks correspond to the solutions of the spectral functions calculated by the 2D $e$-ph coupling model, which is formulated on the basis of the three-dimensional $e$-ph coupling model by Moser $et$ $al$. \cite{Moser2013}. $A^{(0)}$ is given by  $A^{(0)}(E,\bm{k}) \propto f_{\text{FD}}(\epsilon_{\bm{k}}) \cdot |\text{Im} \Sigma |/(|E- \epsilon_{\bm{k}} -\text{Re} \Sigma  |^2  + |\text{Im} \Sigma |^2 ) $, where  $\text{Re} \Sigma$  and $\text{Im} \Sigma$   are the real and imaginary parts of the self-energy $\Sigma$  and $f_{\text{FD}}(\epsilon_{\bm{k}} )$  is the Fermi-Dirac (FD) distribution function.  $\text{Im} \Sigma$  of the SrTiO$_3$ surface was obtained from the ARPES intensity plot by using $2|\text{Im} \Sigma |=(\partial \epsilon_{\bm{k}} / \partial k )\cdot \Delta k $ .  On the other hand,  the spectral function of the $e$-ph satellite state $A^{(1)}(E,\bm{k}) $ is given by
\begin{eqnarray}
A^{(1)}(E,\bm{k}) & \propto &  \sum_{k'}|\gamma_{\text{2D}}(\bm{q}) |^2 f_{\text{FD}}(\epsilon_{\bm{k}'} ) \delta(E -\epsilon_{\bm{k}'} + \hbar\omega_{\text{ph}} )  \nonumber\\
&& \cdot  \delta(\bm{k} -\bm{k}' + \bm{q} ) \Theta(E-E_1 +\hbar\omega_{\text{ph}}),
\end{eqnarray}
where $\delta(x)$ and $\Theta(x)$ are the delta function and the Heaviside step function, respectively. 
In the 2D $e$-ph coupling model, an electron in the first subband with the momentum $\bm{k}'$  and the energy $ \epsilon_{\bm{k}'}=\hbar^2 |\bm{k}'|^2/(2m^*) + E_1$  loses its energy and momentum by an excitation of an LO phonon having a momentum $\bm{q}$ and an energy $\hbar\omega_{\text{ph}}$. Here, we neglected the momentum dependence of the LO phonon energies for simplicity and used the value of  $\hbar\omega_{\text{ph}}=99$ meV for the SrTiO$_3$ surface \cite{Klimin2012}.
 The coupling strength of the 2D electrons and the LO phonons is in a relation of $|\gamma_{\text{2D}}(\bm{q}) |^2\propto 1/|\bm{q}| $ \cite{Peeters1987,sup}.  The $A^{(l>1)}$ components [Fig. 3(b)] are the multiple phonon-loss structures described by a Franck-Condon model and are separated by $\hbar\omega_{\text{ph}}$.

As shown in Fig. 3(c), the spectral weight of $A^{(0)}+ A^{(1)}$ well reproduces the ARPES intensity distribution, supporting the validity of the 2D $e$-ph coupling model.  Fig. 3(d) illustrates a schematic drawing of the energy loss process of a 2D electron by LO phonon excitations.

The same lineshape analysis was carried out for the EDCs of the 2DQP band on the H-dosed ZnO(10$\bar{1}$0) surface, and the results with $\hbar\omega_{\text{ph}}=72$ meV for the ZnO surface \cite{Butkhuzi1998} are shown in Figs. 4(a) and 4(b).  The peak-dip-hump structure in the EDC at $k_x = k_{\text{F}}$ is composed of the $A^{(0)}$ and $A^{(1)}$ peaks, and the long tail at $> E_{\text{F}} - 100$ meV arises from the  $A^{(l> 1)}$ phonon satellite peaks. Absence of the peak-dip-hump structure at $k_x = 0$ nm$^{-1}$ is due to a decrease in an energy separation between the $A^{(0)}$ and $A^{(1)}$ peaks as well as a relative increase in the contribution of the phonon satellite peaks. 
Figures 4(c-f) show the calculated spectral weight distributions of $A^{(0)}$, $A^{(1)}$, and $A^{(0)}+ A^{(1)}$  as well as the measured ARPES intensity distribution.  The spectral weighs of $A^{(0)}$ and $A^{(1)}$ merge at $k_x < 0.5$ nm$^{-1}$, and this distribution captures the features of the experimental data quite reasonably.

\begin{figure}
\centering
\includegraphics[width=7.5cm]{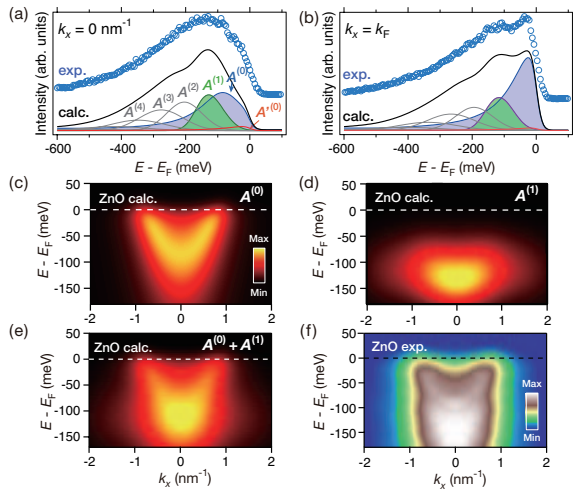}
\caption{ (color online). (a,b) The EDCs (circles) and calculated spectral functions at $k_x = 0$ nm$^{-1}$ (a) and $k_{\text{F}}$ (b) at the H-dosed ZnO(10$\bar{1}$0) surface taken at $h\nu = 65$ eV at 13 K. (c-e) The calculated spectral functions of $A^{(0)}$ (c), $A^{(1)}$ (d), and $A^{(0)}+ A^{(1)}$ (e). (f) The ARPES intensity map (bottom right), which is symmetrized with respect to the center line ($k_x = 0$ nm$^{-1}$) and smoothed. }
\label{fig4}
\end{figure}


The present ARPES study reveals that the broad feature of the 2DES on the H-dosed ZnO(10$\bar{1}$0) surface should be derived from the contribution of the LO phonon satellite structures.  A similar broad feature has also been observed on the metallized ZnO(000$\bar{1}$) surface \cite{Piper2010,Ozawa2011}.  Thus, the $s$-orbital 2D electrons on the ZnO surfaces couple to the LO phonon irrespective of the surface orientation. 
 The $e$-ph interaction and the resultant formation of the spectral tail are regarded as one of the characteristic features of the correlated electron systems \cite{Meevasana2007,Iwasawa2012}, such as the 2DESs on SrTiO$_3$(001) (Fig. 3). These correlated electrons are often referred as an electron liquid \cite{Meevasana2011,Sawatzky1989}. The anomalously large contribution of the phonon satellites in the 2DESs on the ZnO surfaces, therefore, implies that the 2D electrons accumulated on the ZnO surfaces behave like an electron liquid rather than an electron gas.

Finally, we discuss the possible reason why the $e$-ph coupling is prominent for the 2DES on the ZnO surface but not for the surfaces of In$_2$O$_3$ and CdO, although the $s$ electrons contribute to all these 2DESs. The 2D $e$-ph coupling strength is characterized by a dimensionless Fr$\ddot{\text{o}}$hlich electron-phonon coupling constant \cite{Peeters1987},  $\alpha $, which can be derived either from the ratio of the mass of the phonon-dressed electron to that of the bare band mass or from the intensity ratio of the 2D $e$-ph coupling to the quasiparticle states \cite{sup}.  In the case of the 2DES on the H-dosed ZnO(10$\bar{1}$0) surface, the former relation gives $\alpha=0.34 $, while the latter gives $\alpha=0.30 $. On the other hand, negligibly small coupling constants for the 2DESs on In$_2$O$_3$(111) and CdO(100) are predicted by the mass enhancement analysis \cite{sup}. Since the $e$-ph coupling strength tends to be suppressed when the electron density is high because of an efficient electronic screening \cite{Eger1979}, the absence of the phonon satellite structures in these surfaces can be attributed to the higher densities of the 2D electrons. Actually, the densities of the accumulated electrons on In$_2$O$_3$(111) and CdO(100) are $4.2 \times 10^{13}$ cm$^{-2}$ \cite{Zhang2013} and $8 \times 10^{13}$ cm$^{-2}$ \cite{Piper2008}, respectively, and are larger than that on H-dosed ZnO(10$\bar{1}$0) ($0.94 \times 10^{13}$   cm$^{-2}$). In the case of the 2DES on SrTiO$_3$(001), the mass enhancement is absent when the density exceeds $9 \times 10^{13}$ cm$^{-2}$ \cite{Wang2016}. Thus, the electronic screening effect is expected to be more efficient for the $s$ electrons than the $d$ electrons.
From the Fr$\ddot{\text{o}}$hlich $e$-ph coupling constant of a bulk ZnO crystal ($ 0.95$ \cite{Sezen2015}), the 2D limit of the $e$-ph coupling constant is evaluated to be $\alpha \sim$ 0.40 \cite{sup}. Compared to this $\alpha$, the coupling constant obtained at the H-dosed ZnO(10$\bar{1}$0) surface, $\alpha=0.30-0.34$, is slightly suppressed. This suppression is most probably due to the electronic screening. The suppression of the $e$-ph coupling is critically important to enhance the carrier mobilities and the sharpness of the near-UV photoluminescence spectrum \cite{Hauschild2006}.
Therefore, our findings of the reduction of the $e$-ph coupling open up the possibilities that the creation of 2DES enhance the functionalities of, e.g., the  transparent electronic devices \cite{Minami2005} and optoelectronic devices \cite{Hauschild2006}. 

In conclusion, the precise peak and hump structures of the 2DES on the H-dosed ZnO(10$\bar{1}$0) surface have been successfully obtained by ARPES and analyzed with the spectral functions of the 2D $e$-ph coupling model. We found that the $s$-electron derived 2DES is quantitatively described by the model and the 2D electrons at the ZnO surface form the electron liquid. The 2D $e$-ph coupling constant was found to be larger than those expected on the In$_2$O$_3$ and CdO surfaces but smaller than the 2D limit for the bulk ZnO crystal. Our determination of the $e$-ph coupling reveals nature of the 2D electronic states on the ZnO surface and provides technical information for designing novel electronic devices. \\

The authors would like to thank T. Sakurai for the help with the Hall measurements. We gratefully acknowledge A. Fujimori for their valuable discussions. E.F.S. acknowledges financial support from the JSPS postdoctoral fellowship for overseas researchers and the Alexander von Humboldt Foundation (Grant No. P13783). This work was supported by a Grant-in-Aid for Scientific Research (Grant No. 21560695) from MEXT in Japan. This work was carried out by the joint research in the Synchrotron Radiation Research Organization and the Institute for Solid State Physics, the University of Tokyo (Proposal Nos. 2012B7401, 2013A7401, 2013B7401). The synchrotron radiation experiments were performed with the approval of HSRC (Proposal No. 12-B-2). The work at KEK-PF was performed under the approval of the Program Advisory Committee (Proposals No. 2013S2-002) at the Institute of Materials Structure Science, KEK.



\begin{thebibliography}{99}
{

\bibitem{Ohtomo2004} A. Ohtomo and H. Y. Hwang, Nature \textbf{427}, 423 (2004).

\bibitem{Piper2010} L. Piper, A. Preston, A. Fedorov, S. Cho, A. DeMasi, and K. Smith, Phys. Rev. B \textbf{81}, 233305 (2010).

\bibitem{Ozawa2011} K. Ozawa and K. Mase, Phys. Rev. B \textbf{83}, 125406 (2011).

\bibitem{Ozawa2010} K. Ozawa and K. Mase, Phys. Status Solidi A \textbf{207}, 277 (2010).

\bibitem{Ozawa2010a} K. Ozawa and K. Mase, Phys. Rev. B \textbf{81}, 205322 (2010).



\bibitem{Zhang2013} K. H. L. Zhang, R. G. Egdell, F. Offi, S. Iacobucci, L. Petaccia, S. Gorovikov, and P. D. C. King, Phys. Rev. Lett. \textbf{110}, 056803 (2013).

\bibitem{Piper2008} L. F. J. Piper $et$ $al$., Phys. Rev. B \textbf{78}, 165127 (2008).

\bibitem{King2010} P. D. C. King, T. Veal, C. McConville, J. Z{\'{u}}{\~{n}}iga-P{\'{e}}rez, V. Mu{\~{n}}oz-Sanjos{\'{e}}, M. Hopkinson, E. Rienks, M. Jensen, and P. Hofmann, Phys. Rev. Lett. \textbf{104}, 256803 (2010).

\bibitem{Santander-Syro2011} F. Santander-Syro $et$ $al$., Nature \textbf{469}, 189 (2011).

\bibitem{Meevasana2011} W. Meevasana, P. D. C. King, R. H. He, S.-K. Mo, M. Hashimoto, A. Tamai, P. Songsiriritthigul, F. Baumberger, and Z.-X. Shen, Nat. Mater. \textbf{10}, 114 (2011).

\bibitem{DAngelo2012} M. $\text{D'Angelo}$, R. Yukawa, K. Ozawa, S. Yamamoto, T. Hirahara, S. Hasegawa, M. G. Silly, F. Sirotti, and I. Matsuda, Phys. Rev. Lett. \textbf{108}, 116802 (2012).

\bibitem{Yukawa2013} R. Yukawa, S. Yamamoto, K. Ozawa, M. $\text{D'Angelo}$, M. Ogawa, M. G. Silly, F.
Sirotti, and I. Matsuda, Phys. Rev. B \textbf{87}, 115314 (2013).


\bibitem{Wang2016} Z. Wang $et$ $al$., Nat. Mater. \textbf{15}, 835 (2016).


\bibitem{Rodel2016} T. C. R{\"{o}}del $et$ $al$., Adv. Mater. \textbf{28}, 1976 (2016).

\bibitem{King2012} P. D. C. King $et$ $al$., Phys. Rev. Lett. \textbf{108}, 117602 (2012).




\bibitem{Rodel2015} T. C. R{\"{o}}del, F. Fortuna, F. Bertran, M. Gabay, M. J. Rozenberg,  A. F. Santander-Syro, and P. Le F{\`{e}}vre, Phys. Rev. B \textbf{92}, 041106 (2015).






\bibitem{Frohlich1950} H. Fr{\"{o}}hlich, H. Pelzer, and S. Zienau, Philos. Mag. \textbf{41}, 221 (1950).

\bibitem{Deinert2015} J.-C. Deinert, O. T. Hofmann, M. Meyer, P. Rinke, and J. St{\"{a}}hler, Phys. Rev. B \textbf{91}, 235313 (2015).

\bibitem{Ozawa2007} K. Ozawa, T. Sato, Y. Oba, and K. Edamoto, J. Phys. Chem. C \textbf{111}, 4256 (2007).

\bibitem{sup} See Supplemental Material [url] for details of the experimental procedure, the H coverage estimation, the Motto-Ioffe-Regel limit, the band bending calculations, the spectral function calculations, and the $e$-ph coupling constants.

\bibitem{SHIMADA2002} K. Shimada, M. Arita, Y. Takeda, H. Fujino, K. Kobayashi, T. Narimura, H. Namatame, and M. Taniguchi, Surf. Rev. Lett. \textbf{09}, 529 (2002).

\bibitem{Hayashi2013} H. Hayashi, K. Shimada, J. Jiang, H. Iwasawa, Y. Aiura, T. Oguchi, H. Namatame, and M. Taniguchi, Phys. Rev. B \textbf{87}, 035140 (2013).

\bibitem{Zhang2011}  P. Zhang, P. Richard, T. Qian, Y.-M. Xu, X. Dai, and H. Ding, Rev. Sci. Instrum. \textbf{82}, 043712 (2011).

\bibitem{Norman1998} M. R. Norman and H. Ding, Phys. Rev. B \textbf{57}, R11089 (1998).

\bibitem{Cuk2005} T. Cuk, D. H. Lu, X. J. Zhou, Z. X. Shen, T. P. Devereaux, and N. Nagaosa, Phys. Status Solidi B \textbf{242}, 11 (2005).

\bibitem{Moser2013} S. Moser $et$ $al$., Phys. Rev. Lett. \textbf{110}, 196403 (2013).

\bibitem{Butkhuzi1998} T. V Butkhuzi, T. G. Chelidze, A. N. Georgobiani, D. L. Jashiashvili, T. G. Khulordava, and B. E. Tsekvava, Phys. Rev. B \textbf{58}, 10692 (1998).

\bibitem{Goldstein1980} Y. Goldstein, A. Many, I. Wagner, and J. Gersten, Surf. Sci. \textbf{98}, 599 (1980).

\bibitem{Eger1979} D. Eger and Y. Goldstein, Phys. Rev. B \textbf{19}, 1089 (1979).

\bibitem{Yukawa2015} R. Yukawa, K. Ozawa, S. Yamamoto, R.-Y. Liu, and I. Matsuda, Surf. Sci. \textbf{641}, 224 (2015).




\bibitem{Klimin2012} S. N. Klimin, J. Tempere, D. van der Marel, and J. T. Devreese, Phys. Rev. B \textbf{86}, 045113 (2012).


\bibitem{Peeters1987}	F. M. Peeters and J. T. Devreese, Phys. Rev. B \textbf{36}, 4442 (1987).


\bibitem{Meevasana2007} W. Meevasana $et$ $al$., Phys. Rev. B \textbf{75}, 174506 (2007).

\bibitem{Iwasawa2012} H. Iwasawa, Y. Yoshida, I. Hase, K. Shimada, H. Namatame, M. Taniguchi, and Y. Aiura, Phys. Rev. Lett. \textbf{109}, 066404 (2012).

\bibitem{Sawatzky1989}  G. A. Sawatzky, Nature \textbf{342}, 480 (1989).

\bibitem{Sezen2015} H. Sezen $et$ $al$., Nat. Commun. \textbf{6}, 6901 (2015).

\bibitem{Hauschild2006} R. Hauschild, H. Priller, M. Decker, J. Br\"{u}ckner, H. Kalt, and C. Klingshirn, Phys. Status Solidi C \textbf{3}, 976 (2006).

\bibitem{Minami2005} T. Minami, Semicond. Sci. Technol. \textbf{20}, S35 (2005).
















}

\end{thebibliography}
\end{document}